\documentclass[reprint,twocolumn]{revtex4}
\usepackage{amsfonts}
\usepackage{amssymb}
\usepackage{amsmath}
\usepackage{hyperref}
\usepackage{latexsym}
\usepackage {graphicx,fleqn,color}
\usepackage{epsfig}

\begin{document}

\title{Origin of the logarithmic correction to the Newton's law in the
presence of a homogeneous gas of wormholes}
\author{A.A. Kirillov}
\author{ E.P. Savelova}
\affiliation{Dubna International University of Nature, Society and Man, Universitetskaya
Str. 19, Dubna, 141980, Russia }
\date{}

\begin{abstract}
We suggest a new scenario in which the Universe starts its
evolution with a fractal topological structure. This structure is
described by a gas of wormholes. It is shown that the polarization
of such a gas in external fields possesses a spatial dispersion,
which results in a modification of the Newton's law. The
dependence on scales is determined by the distribution of
distances between throat ends. The observed in galaxies
logarithmic correction confirms that the distribution has fractal
properties. We also discus the possibility of restoring such a
distribution from observations.
\end{abstract}

\maketitle

\section{Fractal topological structure as initial conditions}

The most important problem of the modern astrophysics is the nature of dark
matter (DM). While particle physics suggests too many speculative
candidates, LHC does not provide even any hint for the existence of such
dark particles. Moreover, the standard model in particle physics needs not
their existence at all. This make us think that something wrong with our
interpretation of the picture of the Universe observed.

It turns out that lattice quantum gravity, has already suggested a
satisfactory principle answer to the question what DM should be.
The answer lies in the very complex spacetime foam picture which
exists at very small (sub-Planckian) scales and which possesses
fractal properties. Indeed, the first indication that at very
small scales spacetime has a universal fractal structure came from
2d-quantum gravity \cite{KPZ}. This result was obtained both,
analytically and by means of numerical simulations and it has been
developed in many papers, e.g., see \cite{AJW95} and references
therein. Moreover, numerical simulations demonstrate that the
fractal behavior retains in 4d lattice quantum gravity as well.
And indeed, the spectral dimension for nonperturbative quantum
gravity defined via Euclidean dynamical triangulations was
calculated in \cite{AJL05}. It turns out that it runs from a value
of D= 3/2 at short distance to D = 4 at large distance scales.

The fractal structure of the spacetime foam can be described as
follows. Consider any point $x$ in space, fix a geodesic distance
$R$, and consider the value of volume $\Omega _{x}(R)$ which gets
within the ball $d(x,y)\leq R $. \ The homogeneity of the foam
means that $<\Omega _{x}(R)>=\Omega (R)$, i.e., does not depends
on the position of the starting point for geodesic lines. Then the
scaling $\Omega (R)\varpropto R^{D_{H}}$ defines the Hausdorff
dimension of space $D_{H}$. By the construction it is clear that
in 4d gravity $D_{H}\leq 4$. Indeed, the starting point $x$
together with the omnidirectional jet of geodesic lines define the
extrapolating reference system (exactly as we use in astrophysics)
for which the coordinate volume scales with $D=4$. The same
behavior works for simple topology spaces. However, in the
presence of wormholes for sufficiently big distances $R$ some of
geodesics return and start to cover the same (already covered)
physical region $\Omega (R)$. Therefore, for sufficiently complex
topologies
(foam) the dimension is always less than 4 (i.e., we may state that $D_{H}<4$%
).

It is remarkable that we see on the sky exactly such a picture.
Indeed, the light is too scattered upon propagating through the
wormhole throats and it forms the diffused background \cite{K02}.
Therefore, in the homogeneous Universe galaxies are good tracers
for the actual (physical) volume which exactly demonstrate the
fractal behavior $N(R)\varpropto R^{D}$ with \thinspace $D\approx
2$ up to distances $R\sim 200Mpc$ \cite{P87}. In particular, this
reflects the long standing puzzle of missing baryons. The
nontrivial complex topology leads to the modification of Newton's
law \cite{KS07} which is interpreted as the dark matter phenomenon
and which perfectly fits the observations \cite{K06}.

Thus, we see that lattice quantum gravity suggests us a new cosmological
scenario. The inflationary stage in the past has enormously stretched all
scales and the fractal quantum spacetime foam structure has been tempered.
We may say that together with metric perturbations generated from quantum
fluctuations the inflation generates topological perturbations from quantum
spacetime foam as well. Such a structure represents a homogeneous and
isotropic space filled with a gas of wormholes which represents the initial
conditions for the standard cosmological model. In other words, such a space
is homogeneous, while the topological structure is not (its mean value is
homogeneous and isotropic but possesses fluctuations).

Expecting the question on the stability of such wormholes we point
out to the well known fact that the nontrivial topological
structure (the gas of wormholes) is consistent with the
homogeneity of space. Indeed, such a structure can be realized as
space of a constant negative curvature (by means of the cut and
paste technique in the Lobachevsky space). In the three dimensions
this however assumes that the simplest wormhole has throat
sections in the form of a torus or a sphere with a single handle,
since the sphere does not possess the metric of a constant
negative curvature (at least one point will include a
singularity). If the handle on such a sphere is sufficiently
small, it will look almost as a sphere (e.g., as we know the
simplest model of a horse is given by the spherically symmetric
horse in vacuum; In this respect all our attempts to work with
spherically symmetric wormholes look like this and they may have
in the first place the methodological interest). Nevertheless,
upon averaging out over orientations of the torus the spherical
symmetry restores and we may expect that spherical wormholes
reproduce correctly some basic features. We present the simplest
exact example of the constructing of a stable wormhole in the
appendix.

In general, spacetime foam assumes that entrances into wormholes
may be separated by time-like and space-like distances equally.
However it is clear that due to rapid inflationary process the
time-like separation does not survive (it remains of the order of
$\ell _{Pl}$) while space-like distances have enormously change
$\propto e^{H\Delta t}$, where $\Delta t$ is the duration of the
inflationary stage. According to \cite{AJL05} the volume of such a
space scales as $V(R)\varpropto R^{d_{H}}$ with the Hausdorff
dimension between $1<d_{H}\leq 3$. At very small and very large
scales (maybe even outside the Hubble radius) $d_{H}=3$, while on
intermediate scales the dimension changes.

The difference $d_{H}-3\neq 0$ we observe now as the dark matter phenomenon.
Indeed, the simplest estimate gives for the Newton's law $F\varpropto \frac{%
GM}{S(R)}$, where $S(R)$ is the surface of the sphere of the radius $R$
which scales as $S(R)\varpropto R^{d_{H}-1}$. We point out that the value $%
d_{H}\simeq 2$ is in agreement with observations in the range $%
5Kpc<R\lesssim 200Mpc$. The value $d_{H}\simeq 1$ gives too strong
gravitational coupling and such a topological structures should
decay (or even be suppressed during the inflation). Indeed, as we
know metric perturbations do not grow during the radiation
dominated stage. However such a behavior works only for Freedman
spaces (flat space, sphere or the Lobachevsky space). In the
presence of a nontrivial topological structures this, in general,
is not true, e.g., see the first investigation in \cite{KS11}
which has reviled an essential (scale-dependent) modification of
the equations for perturbations.

The negative curvature of space looks to be forbidden by the Doppler picks
in $\Delta T/T$ spectrum. However the upper bounds on the curvature are
model dependent (they are based on simple topology spaces). Moreover, the
apparent value of the curvature is somewhat reduces by the ratio $%
V_{ph}(R)/V_{coor}(R)$, where $V_{ph}(R)\sim R^{d_{H}}$ is the
actual or physical volume of space and $V_{coor}(R)\sim R^{3}$ is
the extrapolating coordinate volume. In this estimates we should
take $R=R_H$ ($R_H$ is the Hubble radius). The same ratio
describes also the portion of missing baryons. In other words,
these problems require the further and more careful investigation.

Thus, we see that our Universe can be rather far from the simple
picture we use. It remains to be isotropic and homogeneous but may
possess a rather complex local topological structure. Our
phenomenological description based on the standard $\Lambda CDM$
model works well enough, but meets some small inconsistencies
which permanently enforce us to add some exotic matter fields
(absent in lab experiments) or consider different modifications of
gravity. However it is rather clear that the polarization of
matter fields on the fractal topology (bias or topological
susceptibility) is rich enough and it is capable of explaining all
exotic properties observed.

\section{Homogeneous gas of wormholes}

Consider now the simplest model suggested by us earlier
\cite{KS07} to demonstrate that the topological bias allows to
mimic dark matter phenomena. As it is explained in the previous
section stable wormholes have throat sections in the form of torus
(e.g., see Fig.3). Therefore the model based on spherical
wormholes (which are not stable) has in the first place the
methodological interest. Nevertheless, it contains all basic
general qualitative features (topological permeability). Moreover,
upon averaging over orientations of tori (wormhole throat
sections) the spherical symmetry of wormholes restores and we may
expect that even some quantitative features will be correct.

When the gravitational field is rather weak, the Einstein equations for
perturbations reduce to the standard Newton's law
\begin{equation*}
\frac{1}{a^{2}}\Delta \phi =4\pi G\left( \delta \rho +\frac{3}{c^{2}}\delta
p\right) ,
\end{equation*}%
here $a$ is the scale factor of the Universe, $\delta \rho $ and
$\delta p$ are the mass density and pressure perturbations
respectively, $G$ is the gravitational constant, and $\Delta
=\nabla ^{2}$ is the Laplace operator. Therefore, the behavior of
perturbations is
determined by the Green function%
\begin{equation*}
\Delta G(x,x^{\prime })=4\pi \delta (r-r^{\prime }).
\end{equation*}%
In the simple flat space the Green function is well known
$G_{0}=-1/r$ (or for Fourier transforms $G_{0}=-4\pi /k^{2}$). In
the presence of wormholes due to polarization on throats the true
Green function obeys formally to the same equation but with biased
source (which is the topological bias or susceptibility)
\begin{equation*}
\Delta G(x,x^{\prime })=4\pi \left( \delta (r-r^{\prime })+b(r-r^{\prime
})\right) .
\end{equation*}%
In the case of a homogeneous gas of wormholes such a bias was evaluated
first in \cite{KS07} (see also more general consideration and details in
\cite{S}) and is given by%
\begin{equation*}
b(k)=2n\overline{R}\frac{4\pi }{k^{2}}\left( \nu (k)-\nu (0)\right) ,
\end{equation*}%
where $n$ is the density of wormhole throats, $\overline{R}$ is the mean
value of the throat radius, and $\nu (k)$ is the Fourier transform for the
distribution over the distances between wormhole mouths. It is defined as $%
\nu (X)=\frac{1}{n\overline{R}}\int F(X_{\pm },R)RdR$, where $F$ is the
number density of wormholes in the configuration space (due to homogeneity $%
X=X_{+}-X_{-}$, and $X_{\pm }$ are positions of wormhole entrances in
space). This function is normalized so that $\int \nu (X)d^{3}X=1$. We point
out that in a more general case (non-spherical wormholes) the bias will have
analogous structure with an appropriate redefinition of the dimensional
parameter $2n\overline{R}\rightarrow 1/\ell ^{2}$, though some details may
change. Thus the true Green function will include some correction and look
like
\begin{equation*}
G\left( k\right) =\frac{-4\pi }{k^{2}\left( 1-b(k)\right) }\simeq
\frac{-4\pi
}{k^{2}}\left( 1+2n\overline{R}\frac{4\pi \left( \nu (k)-\nu (0)\right) }{%
k^{2}}\right) .
\end{equation*}%
In order to get the logarithmic correction (observed in galaxies) the
distribution $\nu (k)$ should give the behavior $b(k)\sim k^{\alpha }$ with $%
\alpha \simeq 1$.

Consider now particular examples of the above destribution. Let all
distances between wormhole entrances are of the same value $r_{0}$, then $%
\nu (X)=\left( 4\pi r_{0}^{2}\right) ^{-1}\delta (|X|-r_{0})$ and we find $%
\nu (k)=\int \nu (X)e^{-ikX}d^{3}X=\frac{\sin kr_{0}}{kr_{0}}$, which gives%
\begin{equation*}
b(k)=-2n\overline{R}\frac{4\pi }{k^{2}}\left( 1-\frac{\sin kr_{0}}{kr_{0}}%
\right) .
\end{equation*}%
For small $kr_{0}\ll 1$ we find
\begin{equation*}
b(k)=\frac{4\pi }{3}n\overline{R}r_{0}^{2}\left( -1+\frac{3!}{5!}\left(
kr_{0}\right) ^{2}+...\right) .
\end{equation*}%
The first constant simply renormalizes the gravitational constant, while
next terms define corrections to the Newt0n's law. At first look such a
decomposition represents a rather general situation. Indeed, Consider an
additional distribution over the parameter $r_{0}$ with any probability
density $p(x)$ ($\int_{0}^{\infty }p(x)dx=1$). Then the same decomposition
works for mean values $<\frac{\sin kr_{0}}{kr_{0}}>=\sum \frac{\left(
-1\right) ^{n}}{\left( 2n+1\right) !}<\left( kr_{0}\right) ^{2n}>$ which
means that the above expression works as well with the replacement $%
r_{0}^{2n}\rightarrow <r_{0}^{2n}>$. However this is possible only for
normal (Gaussian) distributions with stable (finite) momenta. In the case of
fractal picture this is not true. Indeed, Consider a particular fractal
distribution with infinite dispersion of the type%
\begin{equation*}
\nu (k)=\exp \left( -A(ik)^{\alpha }-B(-ik)^{\alpha }\right) .
\end{equation*}%
In the case $\alpha <2$ the dispersion is divergent
\begin{equation*}
\sigma =\frac{d^{2}}{dk^{2}}(\ln \nu (k))|_{k=0}\varpropto k^{\alpha
-2}\rightarrow \infty
\end{equation*}%
and the above decomposition does not work. However all corrections to the
Newton's law can be found from the decomposition of the characteristic
function $\nu (k)$ itself
\begin{equation*}
\nu (k)-1=k^{\alpha }\sum C_{n}k^{\alpha n}.
\end{equation*}%
All the coefficients in the above decomposition reflect deviations from the
standard Newton's law and may be interpreted as the presence of dark matter.
In the first place they should be fixed from observations (the observed
distribution of DM) which allow us to define the actual distribution of
wormholes and restore the true Green function%
\begin{equation*}
G=\frac{-4\pi }{k^{2}}(C+Bk^{\alpha }+...).
\end{equation*}%
For observational needs as an empirical Green function (or generalized
susceptibility) we may suggest the expression%
\begin{equation*}
G_{emp}=\frac{-4\pi }{k^{2}\left( 1+( kr_{0}) ^{-\alpha }\right) }
\end{equation*}%
which at small scales ($kr_{0}\gg 1$) gives the standard Newton's law, while
at large scales $kr_{0}\ll 1$ transforms to the fractal law. We also point
out that the logarithmic correction \ (observed in galaxies) corresponds to
the value $\alpha \approx 1$ and $r_{0}\sim 5Kpc$.
\section{Appendix}
Consider now the cut and paste technique on the Lobachevsky space
which allows to construct  stable wormholes. Consider the upper
complex half-plane, see Fig.1, which gives the model of 2D
Lobachevsky space.
\begin{figure}[tph]
\centering\leavevmode\epsfysize=4.2cm \epsfbox{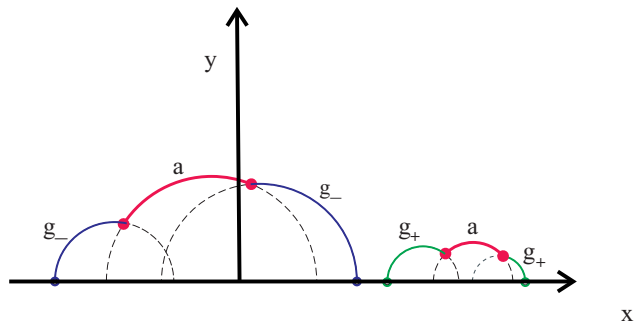}
\newline \caption[fig1]{The upper complex half-plane. $g_{\pm}$ are parts
of geodesic lines which end at the absolute.
$a$ is the equal parts of the two geodesics, which upon gluing
along $g_{\pm}$ transform to the closed geodesic lines as shown on
Fig.2.} \label{fig1}
\end{figure}
The metric on the half-plane has the form
$dl^{2}=\frac{1}{y^{2}}\left( dx^{2}+dy^{2}\right) .$ The absolute
(infinity) is the axis $Ox$ ($y=0$ or $y\rightarrow \infty$).
Geodesic lines are half circles with centers on the absolute, or
perpendicular to the absolute rays. Making the cut along pies-wise
geodesic lines as shown on Fig.1 (solid lines) and gluing along
identical geodesics  $g_{\pm }$ we obtain the space of a constant
negative curvature with two identical closed geodesic boundary
lines $a$ as shown on Fig.2a. Again gluing along closed geodesic
circles $a$ we get the Lobachevsky space with a handle on it, as
shown on Fig.2b.
\begin{figure}[tph]
\centering\leavevmode\epsfysize=8.5cm \epsfbox{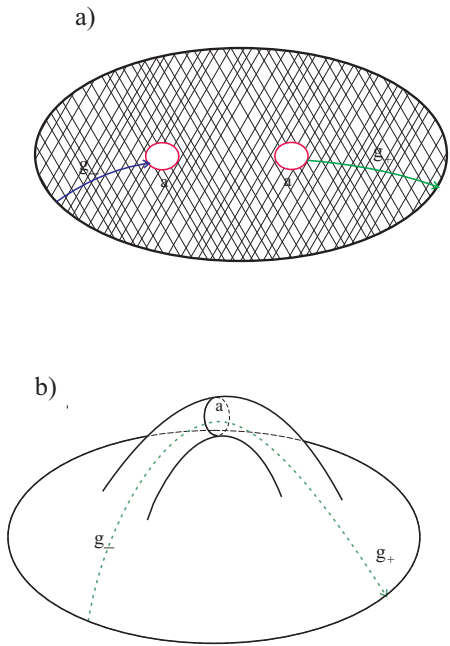}
\newline \caption[fig1]{a) Upon gluing along $g_{\pm }$ we get two identical closed geodesics $a$.
Internal region of the circles $a$ is removed. b) Upon gluing
along $a$ we get the handle in the Lobachevsky plane. Geodesics
$g_{\pm }$ transform into a single continuous geodesic line.}
\label{Fig2}
\end{figure}
It is important that such a space is the space of the constant
curvature. It is also clear that repeating such a procedure we may
insert an arbitrary number of handles.

Now using the axial symmetry of Lobachevsky space we add to the
above 2d wormhole an angle $0\leq \phi <2\pi $, i.e., rotating the
above construction around the axis $Oz$ as shown on Fig.3, we get
the simplest stable 3d wormhole. \begin{figure}[tph]
\centering\leavevmode\epsfysize=4.2cm\epsfbox{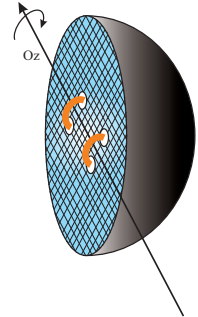}
\newline\caption[fig1]{The ball is the 3D Lobachavsky space. Rotation of its section
around the axis $Oz$ transforms the geodesic circles $a$ on Fig.2a
into the tori. Insides of tori are removed. Upon gluing by the
surfaces of the tori we get the simplest stable wormhole (handle)
in 3D Lobachevsky space.} \label{fig3}
\end{figure}
We repeat again that all these constructions represent spaces with
a constant negative curvature. Their subsequent cosmological
evolution is
governed by the Freedman equations (i.e., the metric takes the form $%
ds^{2}=dt^{2}-a^{2}(t)dl^{2}$, where $dl^{2}$ corresponds to the
Lobachevsky space with a set of wormholes described above).

\end{document}